\documentclass[journal=jacsat,manuscript=article,layout=preprint]{achemso}

\usepackage{graphicx}
\usepackage{dcolumn}
\usepackage{bm}
\usepackage{hyperref}

\usepackage[version=3]{mhchem} 
\usepackage{textcomp}
\usepackage{pslatex }

\newcommand{\onlinecite}[1]{\hspace{-1 ex} \nocite{#1}\citenum{#1}}

\title{Modernist Materials Synthesis: Finding Thermodynamic Shortcuts with Hyperdimensional Chemistry}

\author{James R. Neilson}
\affiliation{Department of Chemistry, Colorado State University, Fort Collins, CO}
\alsoaffiliation{School of Advanced Materials Discovery, Colorado State University, Fort Collins, CO}
\email{james.neilson@colostate.edu}

\author{Matthew J. McDermott}
\affiliation{Materials Sciences Division, Lawrence Berkeley National Laboratory, Berkeley, CA}
\alsoaffiliation{Department of Materials Science and Engineering, University of California, Berkeley, CA}

\author{Kristin A. Persson}
\affiliation{Molecular Foundry, Lawrence Berkeley National Laboratory, Berkeley, CA}
\alsoaffiliation{Department of Materials Science and Engineering, University of California, Berkeley, CA}
\email{kapersson@lbl.gov}

\date{\today}

\begin{document}


\begin{abstract}
Synthesis remains a challenge for advancing materials science. A key focus of this challenge is how to enable selective synthesis, particularly as it pertains to metastable materials.  This perspective addresses the question: how can ``spectator'' elements, such as those found in double ion exchange (metathesis) reactions, enable selective materials synthesis? By observing reaction pathways as they happen (\emph{in situ}) and calculating their energetics using modern computational thermodynamics, we observe transient, crystalline intermediates that suggest that many reactions attain a local thermodynamic equilibrium dictated by local chemical potentials far before achieving a global equilibrium set by the average composition. Using this knowledge, one can thermodynamically ``shortcut'' unfavorable intermediates by including additional elements beyond those of the desired target, providing access to a greater number of intermediates with advantageous energetics and selective phase nucleation. Ultimately, data-driven modeling that unites first-principles approaches with experimental insights will refine the accuracy of emerging predictive retrosynthetic models for complex materials synthesis.
\end{abstract}

\section{Introduction: Broader Context and Key Scientific Questions}\label{sec:intro}

New materials enable new technology. The most significant material in terms of economic impact and quantity, concrete, is enabled by a phase that is metastable at room temperature: \ce{Ca3SiO5}.\cite{west_solid_2014}  This metastability is what allows it to be cured into concrete when mixed with water, \ce{CO2}, and aggregate.  As such, \ce{Ca3SiO5} is synthesized at high temperatures and then quenched rapidly to room temperature to avoid equilibration. Myriad materials predicted to offer interesting functional electronic properties are metastable and not synthesizable under ambient or standard laboratory conditions. Key examples are the successfully predicted high-temperature superconductors, \ce{FeB4}\cite{gou_discovery_2013} and \ce{LaH10},\cite{Drozdov_superconductivity_2019} that were subsequently synthesized at extremely high pressures. Meanwhile, many other functional materials have been predicted to be stable at ambient conditions.\cite{jain_computational_2016} For example, the  battery electrode \ce{Li3FePO4CO3} evaded traditional synthesis and required alternative hydrothermal and ion exchange chemistry for synthesis,\cite{chen_carbonophosphates_2012} thus demanding an improved understanding of synthesis itself.  

\begin{figure}[ht]
    \centering
    \includegraphics[width=3.25in]{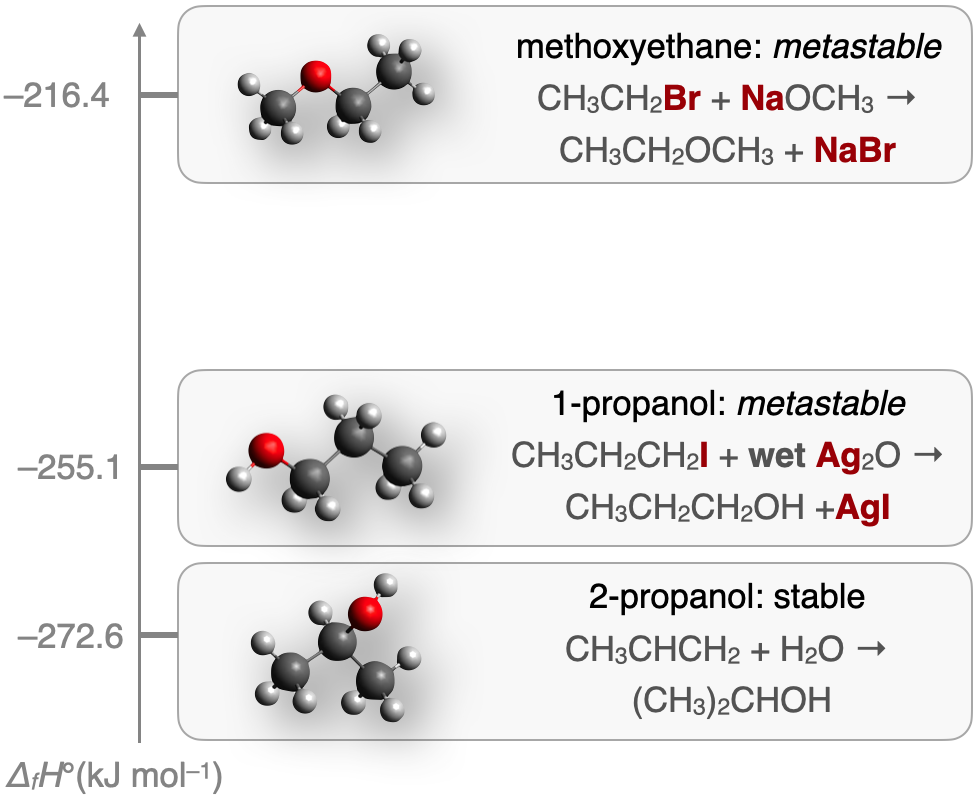}
    \caption{Standard enthalpies of formation of different isomers of \ce{C3H8O}: 2-propanol, 1-propanol, and methoxyethane.\cite{lide_crc_2020}  Also indicated are balanced reactions that include additional species needed for selective reactions to produce the metastable isomers 1-propanol and methoxyethane.}
    \label{fig:isopropanol}
\end{figure}

Accurate prediction of reaction kinetics from first principles requires knowledge of the complex potential energy landscape defined by all atoms in materials, as well as their defects, surfaces, and interfaces.  Therefore, if we assume that reaction kinetics are too challenging to predict accurately for an arbitrary reaction, then how do we take advantage of what we can calculate -- thermodynamics -- to predict useful synthesis reactions?  A key enabling hypothesis explored in this perspective is that reactions between solids often achieve \textit{local} thermodynamic equilibria at interfaces. Several approaches can predict phase equilibria under different intensive and experimentally controllable thermodynamic variables, such as temperature and pressure.\cite{bartel_review_2022} However, we have a relatively poor understanding of how intensive \emph{chemical} variables (e.g., chemical potential, $\mu$) influence the formation of solids, especially when we desire materials that are characterized as metastable.\cite{sun_thermodynamic_2016}  As a thought experiment illustrated in Figure~\ref{fig:isopropanol}, consider the stability and synthesis reactions for various isomers of molecular \ce{C3H8O}.  To make the ground state 2-propanol, one can combine the essential ingredients for a stoichiometric reaction, \ce{CH3CHCH2 + H2O}.  The preparation of other isomers is facilitated by the inclusion of additional elements (e.g., Ag and I in the case of 1-propanol; Na and Br in the case of methoxyethane). We call these additional elements ``spectators'' because they do not appear in the final product. In reality, however, they have a profound influence over reaction selectivity.   

In the context of molecular chemistry, we think about those elements as facilitating specific bond exchanges, since the transformations can occur one bond or one individual molecule at a time.  This contrasts the formation of crystalline solids, during which many bonds exchange in concert.  As such, it is not so straightforward to imagine how selective changes in the chemical formula or chemical structure can manifest at the atomistic scale.  Yet, in some cases, specific elements can be selectively inserted and extracted in topochemical reactions,\cite{england_ion-exchange_1983} like those enabling the lithium-ion battery chemistry of our modern personal electronics era.\cite{mizushima_lixcoo2_1981}  There, the lithium chemical potential can be changed electrochemically to provide a selective driving force for lithium intercalation.  What about the other elements?  How can the presence (or absence) of some elements and different chemical potentials be used for selective synthesis?  

Double ion exchange reactions, also known as ``metathesis'' reactions,\cite{wustrow_metathesis_2022} provide an avenue to address these questions, where the addition of non-essential elements influences the local elemental chemical potentials and the reaction progress.  In the generic reaction scheme to make a target compound $AB$, the simplest reaction could be written as \ce{$A$ + $B$ -> $AB$}. In a metathesis reaction, \ce{$AC$ + $BD$ -> $AB$ + $CD$}, the addition of $C$ and $D$ in the reaction modify the chemical potentials of the species $A$ and $B$. This is because the compounds $AC$ and $BD$ are stable over different ranges of chemical potentials than $A$ and $B$.  Unfortunately, it has remained challenging to predict which species $C$ and $D$ are most suitable for synthesizing an arbitrary target $AB$, as it has historically required knowledge of the full experimental thermodynamic landscape.\cite{yokokawa_generalized_1999}  Today, modern thermodynamic databases fueled by accurate calculation of phase energies using density functional theory (e.g., Materials Project,\cite{jain_commentary_2013} OQMD,\cite{saal_materials_2013} aflowlib\cite{curtarolo_aflow_2012}), have since provided the needed infrastructure for enabling \emph{chemical} control of solid synthesis, as elaborated below.  Furthermore, modern analytical methods, such as synchrotron X-ray and time-of-flight neutron scattering methods, now allow us to interrogate reaction progress as it happens in a so-called ``panoramic synthesis.'' \cite{shoemaker_situ_2014}  By ``cooking and looking,'' we test the thermodynamic hypothesis and can identify the breaking points of the model to move toward a modernist paradigm of predictive retrosynthesis, a ``holy grail'' of computational chemistry.\cite{kovnir2021predictive}

\section{Background: Existing approaches for synthesis prediction}\label{sec:bkgrd}

Two overarching hypotheses allow us to understand and predict the progress of solid state reactions: (a) reactions tend to be dictated by the interfaces between \emph{pairs} of reactants \cite{miura_observing_2021} and (b) the total composition of the reacting system is locally undefined at the interface (i.e., the interface does not ``see'' how far the reactants extend). Several different studies \cite{richards_interface_2016, bianchini_interplay_2020,  aykol_rational_2021} have incorporated one or both of these concepts to deliver new insights.  

\begin{figure}[ht!]
    \centering
    \includegraphics[width=4.in]{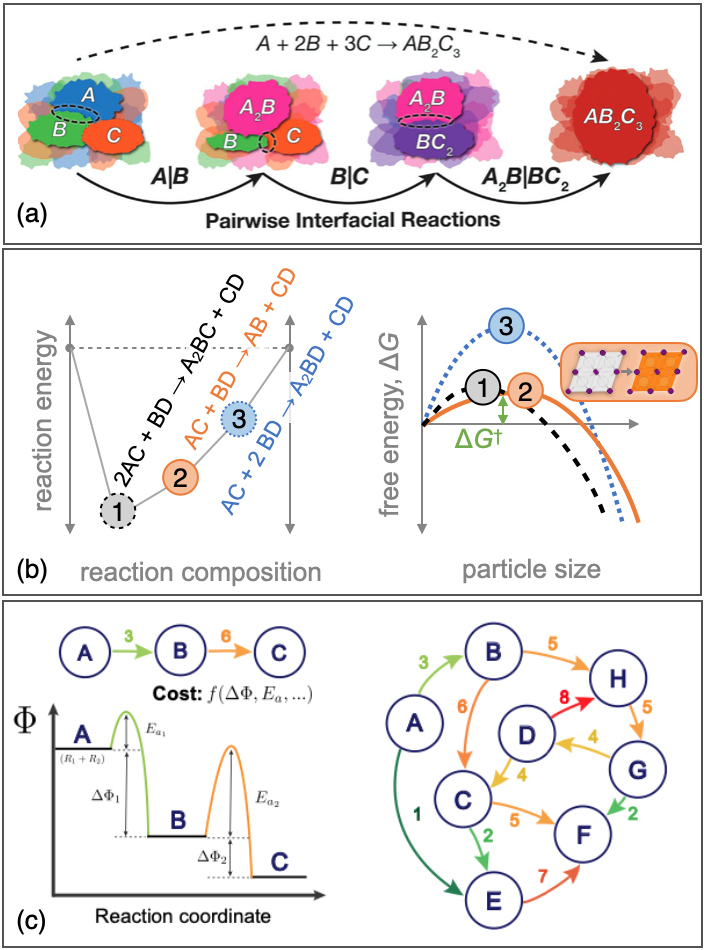}
    \caption{
    Approaches for predicting and understanding materials synthesis using local thermodynamics.
    (a) Pairwise reactions at interfaces tend to represent the pathway of complex reactions involving solids.  Reproduced with permission from Ref.~\cite{miura_observing_2021}. 
    (b) For a given set of phases, one can compute the minimum (most negative) possible reaction energy across a series of compositions.\cite{richards_interface_2016}  A central hypothesis is that the reaction with the most negative free energy, unconstrained by the overall composition, will be the first to occur at the interface (Reaction 1), as the nucleation barrier depends strongly on the reaction energy (Eqn.~\ref{eqn:cnt}).  However, if two phases have a high degree of structural similarity, one can hypothesize that the surface energy contribution of the nucleating phase will be low (e.g., topochemistry) and also result in a low nucleation barrier, $\Delta G^\dagger$ (Reaction 2).\cite{aykol_rational_2021} 
    (c) Schematic reaction coordinate diagram for a simple serial reaction pathway is represented on the left. Actual solid-state reactions take place by multiple, parallel steps and can be better modeled with a reaction network, as shown on the right.  Reproduced with permission from Ref.~\cite{mcdermott_graph-based_2021}.  
    }
    \label{fig:approaches}
\end{figure}

Pairwise reactions have been observed directly with \emph{in situ} electron microscopy during the formation of \ce{YBa2Cu3O_{6+$x$}}, thus enabling understanding of correlated changes in bulk phase fractions determined by \emph{in situ} X-ray diffraction.\cite{miura_observing_2021} Simply, the pairwise interaction between reactants derives from the geometric constraints of a locally planar interface.  As illustrated in Figure~\ref{fig:approaches}(a), the interface between three solid components has a very limited probability of occurrence (note that it is still possible, however, to have a three-component solid state reaction with an ``open'' component such as a liquid or gas).\cite{schmalzried_chemical_2008} Controlling the nature of these interfaces can influence the reaction products.  For example, engineered planar interfaces in  ``modulated elemental reactants'' \cite{johnson_controlled_1998} and sputtered thin films of reactants\cite{mcauliffe_thin-film_2022} provide a geometric control over which pairwise reactions occur first.  In bulk powder reactions, providing a higher density of interfaces directly improves the reaction progress.\cite{kamm_relative_2022} By focusing on pairwise reactions, it is possible to understand why and how those partners evolve together.

Since the interface cannot ``see'' how much of each reactant there is, the first phase to form may not be an equilibrium phase for the composition of the total system. The uncertainty around first-phase formation has created a long-standing problem brought to technological relevance with the development of Si-based microelectronics, where the understanding of metal silicide formation is crucial. Initial hypotheses proposed that the most competitive phase to form at an interface is the congruently melting phase with the highest melting point next to the lowest-temperature eutectic.\cite{walser_first_1976}  
Nowadays, access to large thermodynamic databases enabled by computational chemistry allows us to calculate the most negative reaction energy for the \emph{compositionally-unconstrained} mixture of the two reactants, considering all available mixtures of products (Figure~\ref{fig:approaches}(b)).\cite{richards_interface_2016}  
In the context of synthesis, the first-phase-to-form often establishes the reaction pathway and the subsequent series of steps, as observed during the formation of \ce{MgZrN2} by avoiding unreactive ZrN \cite{rom_bulk_2021} or during the formation of \ce{NaCoO2} in which the most favorable interfacial reaction between CoO and \ce{Na2O2} yields O3-\ce{NaCoO2}, even when the reaction is deficient in \ce{Na2O2}.\cite{bianchini_interplay_2020}  
Relating to classical nucleation theory, the phase with the lowest (most negative) reaction energy should have the smallest activation barrier. This activation barrier scales as
\begin{equation}
\Delta G^\dagger \propto \frac{ \gamma^3}{
 \Delta G_{rxn}^2},\label{eqn:cnt}
\end{equation}
where $\gamma$ is a term denoting the energy cost of forming a surface of the nucleating phase, $\Delta G_{rxn}$ is the free energy of phase formation per volume, and shape effects are ignored.\cite{dheurle_nucleation_1988} Thus, there is a fundamental relationship that links thermodynamic quantities (reaction energy, surface energy) to the relative rates of nucleation for different competing intermediates, with the assumption that a higher nucleation rate yields a dominant reaction rate.  However, it is challenging to assert that such reactions are truly limited by nucleation when reactions occur in a diffusion-limited regime. Regardless, if one assumes that diffusion is universally limiting, a minimal nucleation barrier can provide an advantage.  A complementary approach for assessing the relative rates of phase formation without explicit calculation of surface energies is to estimate relative reaction barriers from structural self-similarity based on the principle of heterogeneous nucleation. In other words, reactants that are more structurally similar to their products ought to have a lower barrier.\cite{aykol_rational_2021} This is schematically illustrated in Figure~\ref{fig:approaches}(b):  while Reaction (1) has the most negative reaction energy, Reaction (2) may have a comparably low activation barrier for formation due to the structural similarity between precursors and products.  Building on the notion of ``phase selection through nucleation,'' \cite{aykol_rational_2021} these physical principles allow the data-enabled synthetic scientist to start making predictions.  


In modern materials synthesis planning, we identify connections and correlations within the large landscape of databases that link together composition, structure, and thermodynamic potentials.  When we consider a simplistic notion of a reaction coordinate diagram, we can consider the energy balance progressing in a serial and linear fashion, as illustrated in Figure~\ref{fig:approaches}(c).  However, a more realistic representation is to consider the parallel but pairwise reactions with different reactants and  intermediates in a graph-based network, in which each node is a phase (or mixture of phases), and each edge is an individual reaction weighted by its ``cost''. In its most simple form, the cost of a reaction is often a function of its energy, $\Delta G_\text{rxn}$ (Figure~\ref{fig:approaches}(d)).\cite{mcdermott_graph-based_2021}  The modernist era provides us with access to a very large available landscape in this representation, with numerous approaches to weight the edge lengths and algorithms to identify the lowest cost pathways through the network.\cite{wen_chemical_2023} 
With brute-force enumeration of all possible reactions using these large databases, selective reactants have been identified as those with the fewest competing reaction products that also have a favorable reaction energy.\cite{aykol_rational_2021}  Altogether, there is a powerful quiver at our disposal for enabling a new paradigm of materials synthesis.

\section{Phase selection from local chemical potentials}\label{sec:science}

\begin{figure}[ht!]
    \centering
    \includegraphics[width=4.5in]{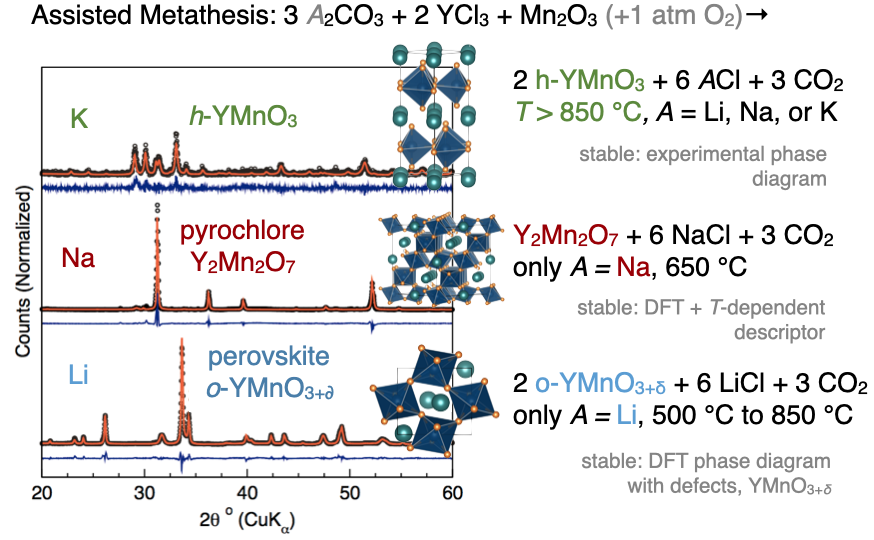}
    \caption{Observed reaction products for different $A$ = Li, Na, vs K in the reaction, \ce{3\ $A$2CO3 + 2\ YCl3 + Mn2O3 + O2\ (\text{1 atm})}: $o$-\ce{YMnO_{3+$\delta$}}, \ce{Y2Mn2O7}, and $h$-\ce{YMnO3},\cite{todd_selective_2019} their crystal structures, and known stability conditions.  Adapted with permission from Ref.~\cite{todd_selective_2019}. Copyright 2019 American Chemical Society}
    \label{fig:metathesis}
\end{figure}

With the conceptual stage set -- unconstrained reservoirs, large thermodynamic databases, cooking-and-looking -- it is now possible to rationalize unusual observations from synthesis studies, such as the use of ``assisted'' metathesis to produce different Y-Mn-O ternary phases.  In the assisted metathesis reaction (cf., Ref.~[\onlinecite{Toberer2004}]), \ce{3\ $A$2CO3 + 2\ YCl3 + Mn2O3 + \text{1 atm}\ O2}, where $A$ is an alkali, each alkali yields different products when the reaction temperature is below 850 \textcelsius{} (see Fig~\ref{fig:metathesis}).  For $A$ = Li, the reaction yields the \emph{perovskite} \ce{YMnO3} polymorph ($o$-\ce{YMnO3}), yet when $A$ = Na, the reaction yields the \emph{pyrochlore} \ce{Y2Mn2O7} polymorph under otherwise identical synthesis conditions.  For $A$ = K, the reaction yields a mixture of phases below $T < 850$\textcelsius{}, and exclusively the \emph{hexagonal} \ce{YMnO3} polymorph for $T \ge 850$\textcelsius{} ($h$-\ce{YMnO3}).  When the reaction temperature exceeds 850 \textcelsius{}, all three alkali species yield $h$-\ce{YMnO3}. These observations raise the questions: (1) which of the three ternary Y-Mn-O phases are stable under what conditions, and (2) how/why does the alkali ``spectator'' ion produce such distinct changes in product formation?

From the preceding experimental phase diagram and other studies of high-temperature ceramic reactions,\cite{fedorova_heterogeneous_2004,smith_mn_2009} $h$-\ce{YMnO3} was reported to be the bulk stable phase above $T=850$ \textcelsius{}.  Synthesis of $o$-\ce{YMnO3} typically requires quite oxidizing conditions, such as 35 kbar ($\approx$3.5 GPa) of oxygen pressure at $T_\text{rxn}$ = 1000 \textcelsius{},\cite{wood_magnetic_1973} or curiously, a lower-temperature reaction at 800 \textcelsius{} using an amorphous precursor formed by the decomposition of citrate salts.\cite{brinks_synthesis_1997}  In the modernist approach, we illustrated that the incorporation of metal vacancies from oxidation (e.g., \ce{Y_{$1-x$}Mn_{$1-y$}O3}, also referred to as \ce{YMnO_{$3+\delta$}}) thermodynamically stabilizes the $o$-\ce{YMnO3} polymorph \cite{todd_defect-accommodating_2020}  in agreement with $T$- and $p$\ce{O2}-dependent metathesis reactions.\cite{todd_yttrium_2019}  \ce{Y2Mn2O7} typically requires rather extreme conditions for synthesis, such as 30 kbar \ce{O2} reactions with \ce{KIO4} at $T=$ 1000 \textcelsius{},\cite{fujinaka_syntheses_1979} or 500 \textcelsius{} hydrothermal reactions with \ce{ClO3^-}.\cite{subramanian_ferromagnetic_1988}  In the modernist approach, with density functional theory calculations in conjunction with $T$-dependent elemental references and a machine-learned expression for vibrational entropy,\cite{bartel_physical_2018} we find that \ce{Y2Mn2O7} is thermodynamically stable below $T\le$ 1000 \textcelsius{} and favored over \ce{YMnO3} when $p$\ce{O2} $\ge$ 0.7 atm.\cite{todd_selectivity_2021}.  It follows that \ce{Y2Mn2O7} should actually be formed in all of these assisted metathesis reactions, given that they are performed at $p$\ce{O2} $\approx$ 0.85 atm. Therefore, the alkali ``spectators'' must somehow influence the reaction pathway to shift the local equilibrium.  It is sensible that a lower-temperature reaction can yield a more oxidized product; however, it is not yet clear why \ce{Li2CO3} and \ce{Na2CO3} precursors arrive at different reaction products even when performed at an identical temperature and $p$\ce{O2}. The fact that the Na-based precursor leads to the more oxidized product, \ce{Y2Mn2O7}, is even more perplexing given that the Na-based reaction is more exothermic and could yield more reducing conditions than the Li-based reaction if the local interface experiences quasi-adiabatic self-heating.


\begin{figure*}[ht!]
    \centering
    \includegraphics[width=6.5in]{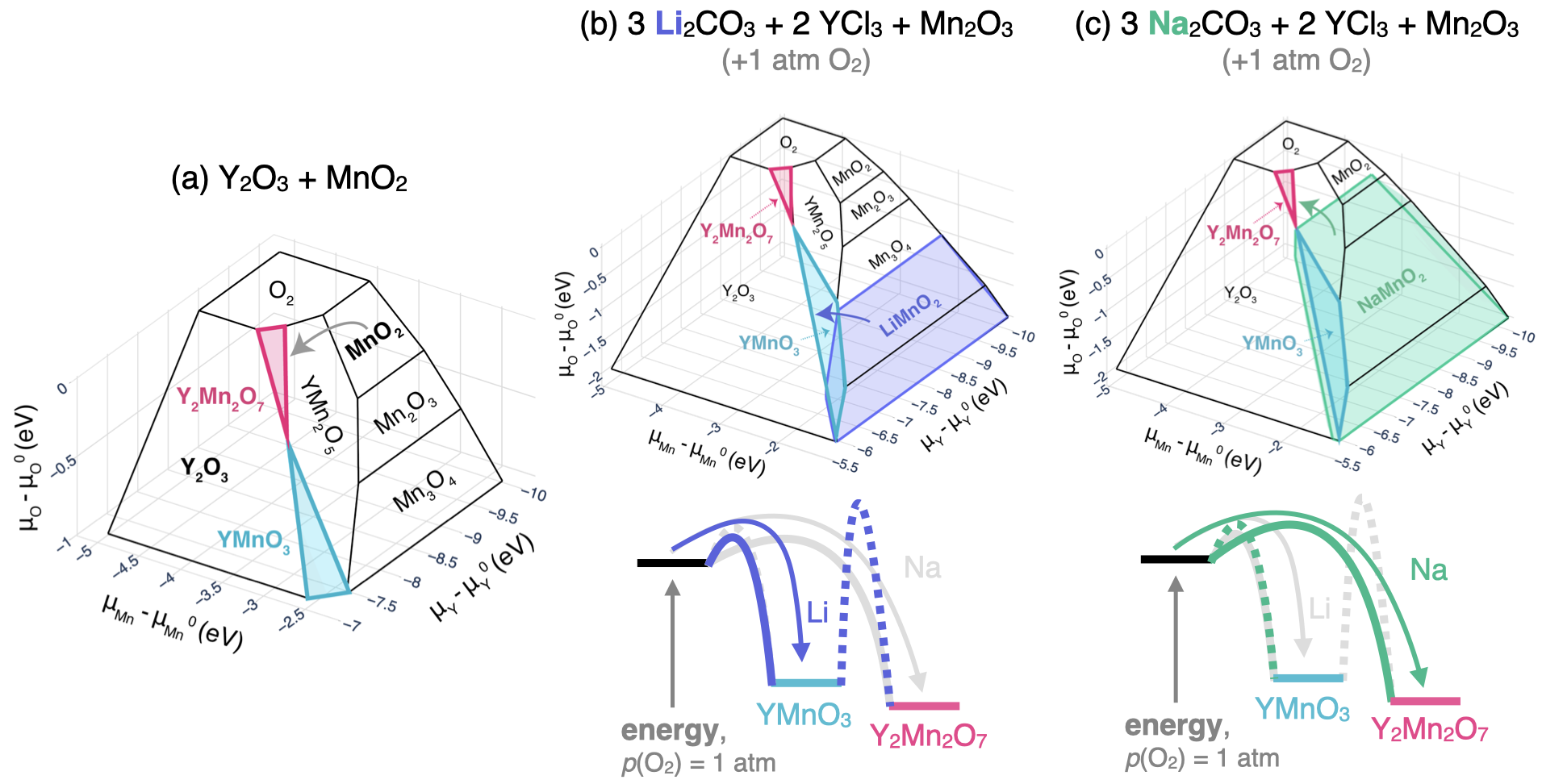}
    \caption{(a) Generalized chemical potential diagram of the Y-Mn-O system. Reaction of \ce{Y2O3 + 2\ MnO2} should result in the initial formation of some amount of \ce{YMn2O5}.  (b) Y-Mn-O chemical potential diagram with a projection of the \ce{LiMnO2} domain.\cite{todd_selectivity_2021} The illustrated overlap between this domain and \ce{YMnO3} suggests a lower effective activation barrier for \ce{LiMnO2} to form \ce{YMnO3}.
    (c) Similarly, the location of the \ce{NaMnO2} overlapping region\cite{todd_selectivity_2021} allows for a lower effective activation barrier in forming \ce{Y2Mn2O7}.  Adapted with permission from Ref.~\cite{todd_selectivity_2021}. Copyright 2021 American Chemical Society}
    \label{fig:mu_map}
\end{figure*}

To understand the interface between reactants, the intensive variable of chemical potential, $\mu$, best describes the local equilibrium between effectively unrestricted reservoirs of reacting elemental species.  All interfaces evolve towards a local equilibrium where atomic chemical potentials equalize across the interface, as the rate of mass transfer is proportional to the chemical potential gradient.  While a large difference in chemical potential can provide a high rate of diffusion (and thus reaction), if the difference is too large, it may also necessitate the formation of another phase that is stable over intermediate ranges of chemical potential.  This fact also explains why synthesis reactions from pure elemental precursors (mixed in the proper ratios) do not always result in the expected products despite possessing the most significant thermodynamic advantage (i.e., the most negative reaction energy possible). The difficulty in synthesizing a target product is especially true when multiple polymorphs may be accessible, such as with the isomers in Figure~\ref{fig:isopropanol}. A thermodynamic explanation is that the elemental precursor and desired target often do not form a stable interface due to competition from other phases that form over intermediate chemical potentials. By adding spectator ions to our elemental precursors, we can modify the chemical potentials of the reacting species such that a stable interface should be formed with our desired target, supporting direct and selective reaction to a product that is stable over a particular range of chemical potentials. The addition of spectator elements also increases the dimensionality of the chemical potential space, allowing one to bypass undesired intermediates. Finally, if the spectator ions together form an energetically favorable byproduct (e.g., an alkali halide), then it is possible to ``re-claim'' some of the reaction driving force that is lost to using a more stable precursor (i.e., one containing bonds with the spectator species).

Mapping the Y-Mn-O phase equilibria in terms of the elemental chemical potentials on a generalized chemical potential diagram (e.g., a 3-dimensional diagram with the axes, $\mu_\text{Y}$, $\mu_\text{Mn}$, and $\mu_\text{O}$, Figure~\ref{fig:mu_map}(a))\cite{yokokawa_generalized_1999} allows us to directly observe which phases exhibit stable interfaces with each other. This map of stable interfaces rationalizes the observation of specific intermediate compounds. For example, a reaction of \ce{Y2O3 + 2 MnO2} or \ce{Y2O3 + Mn2O3} should necessarily produce \ce{YMn2O5} on route to forming \ce{Y2Mn2O7} or \ce{YMnO3}. Indeed, previous work\cite{todd_selective_2019,todd_yttrium_2019} has shown that for $T \le 850 $ \textcelsius{}, the reaction halts at \ce{YMn2O5} and does not yield stoichiometric completion of the reaction which would be the global thermodynamic equilibrium.\cite{fedorova_heterogeneous_2004}  

However, the assisted metathesis reactions proceed through \ce{$A_x$MnO2} and \ce{YOCl} intermediates, as learned from \emph{in situ} synchrotron X-ray diffraction experiments. To visualize this high-dimensional phase equilibria, it is possible to extract lower-dimensional slices of the full $N=6$ high-dimensional phase space, as illustrated by the intersection of the different \ce{$A_x$MnO2} phases onto the Y-Mn-O chemical subspace (Figure~\ref{fig:mu_map}).  Focusing on the role of \ce{LiMnO2} in Figure~\ref{fig:mu_map}(b), it shares a stable interface with \ce{YMnO3} but not \ce{Y2Mn2O7}.  Therefore, to maintain a local equilibrium, the \ce{LiMnO2} intermediate would have to subsequently form \ce{Li2MnO3}, \ce{LiMn2O4}, and ``\ce{Li5Mn7O16}'' (each with distinct crystal structures) to produce  \ce{Y2Mn2O7}.\cite{todd_selectivity_2021}  As the chemistry proceeds at a lower temperature through a defective \ce{Li$_x$MnO2} intermediate, the higher $\mu_\text{O}$ stabilizes the $o$-\ce{YMnO3} polymorph.\cite{todd_defect-accommodating_2020}   In other words, we infer that \ce{LiMnO2} has a lower effective activation barrier to form \ce{YMnO3} than \ce{Y2Mn2O7} because the reaction does not need to proceed through subsequent intermediates. 

In contrast, the assisted metathesis reaction with Na proceeds through \ce{Na$_x$MnO2} intermediates.  Visualized in Figure~\ref{fig:mu_map}(c), the projection of \ce{NaMnO2} into the Y-Mn-O subspace reveals a stable interface with \ce{Y2Mn2O7}.  Hence, the Na-containing reaction exhibits a lower effective activation barrier to nucleate \ce{Y2Mn2O7} and does not necessitate the formation of additional intermediates. In the case of \ce{KMnO2}, it forms stable interfaces with both \ce{YMnO3} and \ce{YMn2O5} and indeed, a mixture of both of these products forms after 24 h reactions at $T_\text{rxn}$ = 850 \textcelsius{}.\cite{todd_selective_2019}  These \emph{in situ} observations, coupled with a map of the high-dimensional chemical space, illustrate that different ``spectator'' elements modify the chemical potentials of the precursors differently, thereby changing which intermediates and/or products are accessible.  This understanding enables a chemistry-based control of phase selection.

\begin{figure}[h!]
    \centering
    \includegraphics[width=3.5in]{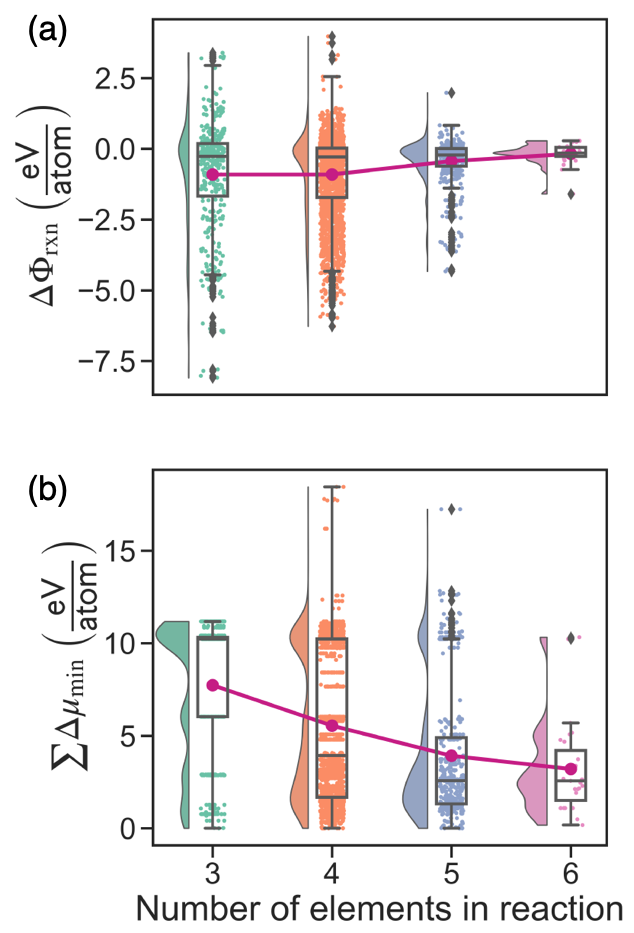}
    \caption{As illustrated for the nearly 3000 enumerated reaction pathways to synthesize \ce{Y2Mn2O7} in the Na-Cl-Y-Mn-O-C chemical system,\cite{todd_selectivity_2021} (a) including additional elements in a reaction tends not to significantly shift the mean grand potential reaction energy, $\Delta \Phi_\text{rxn}$), when examining all enumerated pathways to form \ce{Y2Mn2O7}. (b) However, increasing the number of elements in the system does significantly decrease the total change in chemical potential (e.g., chemical potential ``distance''), thus illustrating the presence of thermodynamic shortcuts. Reprinted with permission from Ref.~\cite{todd_selectivity_2021}. Copyright 2021 American Chemical Society}
    \label{fig:hyperdimension}
\end{figure}

More generally, the addition of non-essential elements into the reaction facilitates the reaction progress with thermodynamics alone.  For example, when considering over 3000 different enumerated reactions to produce \ce{Y2Mn2O7}, the average change in local chemical potentials (i.e., the minimum chemical potential ``distance'' as measured on a chemical potential diagram) decreases systematically with the number of elements while not significantly affecting the mean reaction energy available (Figure~\ref{fig:hyperdimension}). This can be intuited: as you add more ``spectators,'' you increase the likelihood that there will be an intermediate phase that forms stable interfaces between reactant(s) and product(s), or between other intermediates. This also matches geometric intuition -- by increasing the number of dimensions of our chemical potential space, we increase the available number of phase boundaries shared with the desired product(s). The spectator approach is a way of replacing the need to form two or more intermediates with only one, or possibly none at all. This is, in effect, a shortcut through the now larger accessible thermodynamic phase space and a means to affect selectivity.

\section{Outlook}\label{sec:outlook}
When we consider ``hyperdimensional'' chemical reactions -- reactions that include elements beyond those contained in the product -- the reaction design space increases dramatically.  In a previous perspective concerning ``kinetic control'' in materials synthesis,\cite{martinolich_toward_2017} we voiced the challenge of how to selectively control reaction barriers in a general manner.  Here, we effect this notion by using the local chemical potential to minimize the number of phases that need to nucleate.  The chemical control provides thermodynamic shortcuts as a means for ``phase selection through nucleation.''  

Looking forward, a significant challenge is finding other non-trivial reactions that test these principles.  In the synthesis of \ce{YMnO3} phases, it was recognized many years ago that ``kinetic factors'' played a role in reaction outcomes.\cite{giaquinta_structural_1994}  Here we argue that these kinetic factors are intuited by the formation of local chemical equilibrium intermediates that effectively halt the reaction. To explore to what extent the broader term kinetic factors can be cast into a more quantitative framework of local thermodynamics, we need to test our hypothesis with additional systems.  We note that our hypothesis encompasses the notion of ``remnant metastability,''\cite{sun_thermodynamic_2016} such that the phases we call ``metastable'' often form under conditions where that phase is indeed, locally or globally, the most stable phase that can be formed at an interface.  Knobs that can be turned to affect the thermodynamic landscape include e.g. strain, dimensionality, external fields, and chemical potential.\cite{sun_generalized_2021}  Given the use of  databases and natural language processing techniques,\cite{kononova_text-mined_2019,kim_materials_2017,he_similarity_2020,huo_machine-learning_2022}  modernist materials chemists have many tools at their disposal.  

If our goal is to arrive at a fully predictive model for synthesis planning, many fundamental knowledge gaps remain.  Examining thermodynamic variables alone (e.g., reaction energies, minimum changes in chemical potentials), we do not yet understand how to accurately and efficiently include the role of atomistic disorder when assessing energy or diffusion.  In reactivity models using graph networks, the ``kinetic'' barrier imagined to contribute to the distance between nodes is still unknown\cite{wen_chemical_2023}  
Furthermore, we acknowledge that in many real-life cases, there is a true time dependence to the reaction outcome (e.g., the specific 17$\pm$0.5 h reaction required to make superconducting \ce{Bi3O2S3} \cite{phelan_stacking_2013}).  In fact, there is a clear variability in the formation of \ce{Y2Mn2O7}, such that  some reactions, unexpectedly, produce a mixture of \ce{Y2O3} and \ce{Mn2O3} binaries.\cite{todd_selective_2019,wustrow_reaction_2022}  Given the immense complexity of the true potential energy landscape and the time-dependent kinetic processes that influence the movement across that landscape, new modernist approaches beyond reaction-diffusion kinetics \cite{dybkov_reaction_1986}  will enable significant advances. Different communities with different tools and approaches united by fundamental synthesis questions -- be it for a specific application or scientific curiosity -- have made significant strides and still have many exciting discoveries to be made at the next interface.  


\section{Acknowledgements}

This work was supported as part of GENESIS: A Next Generation Synthesis Center, an Energy Frontier Research Center funded by the U.S. Department of Energy, Office of Science, Basic Energy Sciences under Award Number DE-SC0019212. 

\section{Conflict of Interest Statement}

On behalf of all authors, the corresponding author states that there is no conflict of interest.

\section{Data Availability}

The full set of enumerated chemical reactions, as well as all experimental raw data, processed data, data processing scripts, and figure plotting scripts, are available at: \\ \href{https://github.com/GENESISEFRC/y2mn2o7-selectivity}{https://github.com/GENESISEFRC/y2mn2o7-selectivity},  and \\
\href{https://github.com/GENESIS-EFRC/reaction-network}{https://github.com/GENESIS-EFRC/reaction-network}.

\bibliography{SynthesisScience}

\providecommand{\latin}[1]{#1}
\makeatletter
\providecommand{\doi}
  {\begingroup\let\do\@makeother\dospecials
  \catcode`\{=1 \catcode`\}=2 \doi@aux}
\providecommand{\doi@aux}[1]{\endgroup\texttt{#1}}
\makeatother
\providecommand*\mcitethebibliography{\thebibliography}
\csname @ifundefined\endcsname{endmcitethebibliography}
  {\let\endmcitethebibliography\endthebibliography}{}
\begin{mcitethebibliography}{53}
\providecommand*\natexlab[1]{#1}
\providecommand*\mciteSetBstSublistMode[1]{}
\providecommand*\mciteSetBstMaxWidthForm[2]{}
\providecommand*\mciteBstWouldAddEndPuncttrue
  {\def\EndOfBibitem{\unskip.}}
\providecommand*\mciteBstWouldAddEndPunctfalse
  {\let\EndOfBibitem\relax}
\providecommand*\mciteSetBstMidEndSepPunct[3]{}
\providecommand*\mciteSetBstSublistLabelBeginEnd[3]{}
\providecommand*\EndOfBibitem{}
\mciteSetBstSublistMode{f}
\mciteSetBstMaxWidthForm{subitem}{(\alph{mcitesubitemcount})}
\mciteSetBstSublistLabelBeginEnd
  {\mcitemaxwidthsubitemform\space}
  {\relax}
  {\relax}

\bibitem[West(2014)]{west_solid_2014}
West,~A.~R. \emph{Solid state chemistry and its applications}, second edition,
  student edition ed.; Wiley: Chichester, West Sussex, UK, 2014\relax
\mciteBstWouldAddEndPuncttrue
\mciteSetBstMidEndSepPunct{\mcitedefaultmidpunct}
{\mcitedefaultendpunct}{\mcitedefaultseppunct}\relax
\EndOfBibitem
\bibitem[Gou \latin{et~al.}(2013)Gou, Dubrovinskaia, Bykova, Tsirlin,
  Kasinathan, Schnelle, Richter, Merlini, Hanfland, Abakumov, Batuk,
  Van~Tendeloo, Nakajima, Kolmogorov, and Dubrovinsky]{gou_discovery_2013}
Gou,~H.; Dubrovinskaia,~N.; Bykova,~E.; Tsirlin,~A.~A.; Kasinathan,~D.;
  Schnelle,~W.; Richter,~A.; Merlini,~M.; Hanfland,~M.; Abakumov,~A.~M.;
  Batuk,~D.; Van~Tendeloo,~G.; Nakajima,~Y.; Kolmogorov,~A.~N.; Dubrovinsky,~L.
  Discovery of a {Superhard} {Iron} {Tetraboride} {Superconductor}.
  \emph{Physical Review Letters} \textbf{2013}, \emph{111}, 157002\relax
\mciteBstWouldAddEndPuncttrue
\mciteSetBstMidEndSepPunct{\mcitedefaultmidpunct}
{\mcitedefaultendpunct}{\mcitedefaultseppunct}\relax
\EndOfBibitem
\bibitem[Drozdov \latin{et~al.}(2019)Drozdov, Kong, Minkov, Besedin,
  Kuzovnikov, Mozaffari, Balicas, Balakirev, Graf, Prakapenka, Greenberg,
  Knyazev, Tkacz, and Eremets]{Drozdov_superconductivity_2019}
Drozdov,~A.~P.; Kong,~P.~P.; Minkov,~V.~S.; Besedin,~S.~P.; Kuzovnikov,~M.~A.;
  Mozaffari,~S.; Balicas,~L.; Balakirev,~F.~F.; Graf,~D.~E.; Prakapenka,~V.~B.;
  Greenberg,~E.; Knyazev,~D.~A.; Tkacz,~M.; Eremets,~M.~I. Superconductivity at
  250 {K} in lanthanum hydride under high pressures. \emph{Nature}
  \textbf{2019}, \emph{569}, 528--531\relax
\mciteBstWouldAddEndPuncttrue
\mciteSetBstMidEndSepPunct{\mcitedefaultmidpunct}
{\mcitedefaultendpunct}{\mcitedefaultseppunct}\relax
\EndOfBibitem
\bibitem[Jain \latin{et~al.}(2016)Jain, Shin, and
  Persson]{jain_computational_2016}
Jain,~A.; Shin,~Y.; Persson,~K.~A. Computational predictions of energy
  materials using density functional theory. \emph{Nature Reviews Materials}
  \textbf{2016}, \emph{1}, 15004\relax
\mciteBstWouldAddEndPuncttrue
\mciteSetBstMidEndSepPunct{\mcitedefaultmidpunct}
{\mcitedefaultendpunct}{\mcitedefaultseppunct}\relax
\EndOfBibitem
\bibitem[Chen \latin{et~al.}(2012)Chen, Hautier, Jain, Moore, Kang, Doe, Wu,
  Zhu, Tang, and Ceder]{chen_carbonophosphates_2012}
Chen,~H.; Hautier,~G.; Jain,~A.; Moore,~C.; Kang,~B.; Doe,~R.; Wu,~L.; Zhu,~Y.;
  Tang,~Y.; Ceder,~G. Carbonophosphates: {A} {New} {Family} of {Cathode}
  {Materials} for {Li}-{Ion} {Batteries} {Identified} {Computationally}.
  \emph{Chemistry of Materials} \textbf{2012}, \emph{24}, 2009--2016\relax
\mciteBstWouldAddEndPuncttrue
\mciteSetBstMidEndSepPunct{\mcitedefaultmidpunct}
{\mcitedefaultendpunct}{\mcitedefaultseppunct}\relax
\EndOfBibitem
\bibitem[Bartel(2022)]{bartel_review_2022}
Bartel,~C.~J. Review of computational approaches to predict the thermodynamic
  stability of inorganic solids. \emph{Journal of Materials Science}
  \textbf{2022}, \emph{57}, 10475--10498\relax
\mciteBstWouldAddEndPuncttrue
\mciteSetBstMidEndSepPunct{\mcitedefaultmidpunct}
{\mcitedefaultendpunct}{\mcitedefaultseppunct}\relax
\EndOfBibitem
\bibitem[Sun \latin{et~al.}(2016)Sun, Dacek, Ong, Hautier, Jain, Richards,
  Gamst, Persson, and Ceder]{sun_thermodynamic_2016}
Sun,~W.; Dacek,~S.~T.; Ong,~S.~P.; Hautier,~G.; Jain,~A.; Richards,~W.~D.;
  Gamst,~A.~C.; Persson,~K.~A.; Ceder,~G. The thermodynamic scale of inorganic
  crystalline metastability. \emph{Science Advances} \textbf{2016}, \emph{2},
  e1600225\relax
\mciteBstWouldAddEndPuncttrue
\mciteSetBstMidEndSepPunct{\mcitedefaultmidpunct}
{\mcitedefaultendpunct}{\mcitedefaultseppunct}\relax
\EndOfBibitem
\bibitem[England \latin{et~al.}(1983)England, Goodenough, and
  Wiseman]{england_ion-exchange_1983}
England,~W.; Goodenough,~J.; Wiseman,~P. Ion-exchange reactions of mixed
  oxides. \emph{Journal of Solid State Chemistry} \textbf{1983}, \emph{49},
  289--299\relax
\mciteBstWouldAddEndPuncttrue
\mciteSetBstMidEndSepPunct{\mcitedefaultmidpunct}
{\mcitedefaultendpunct}{\mcitedefaultseppunct}\relax
\EndOfBibitem
\bibitem[Mizushima \latin{et~al.}(1981)Mizushima, Jones, Wiseman, and
  Goodenough]{mizushima_lixcoo2_1981}
Mizushima,~K.; Jones,~P.; Wiseman,~P.; Goodenough,~J. {LixCoO2}
  (0{\textless}x$\leq$1): {A} new cathode material for batteries of high energy
  density. \emph{Solid State Ionics} \textbf{1981}, \emph{3-4}, 171--174\relax
\mciteBstWouldAddEndPuncttrue
\mciteSetBstMidEndSepPunct{\mcitedefaultmidpunct}
{\mcitedefaultendpunct}{\mcitedefaultseppunct}\relax
\EndOfBibitem
\bibitem[Lide and Kehiaian(2020)Lide, and Kehiaian]{lide_crc_2020}
Lide,~D.~R.; Kehiaian,~H.~V. \emph{{CRC} {HANDBOOK} of {THERMOPHYSICAL} and
  {THERMOCHEMICAL} {DATA}}, 1st ed.; CRC Press, 2020\relax
\mciteBstWouldAddEndPuncttrue
\mciteSetBstMidEndSepPunct{\mcitedefaultmidpunct}
{\mcitedefaultendpunct}{\mcitedefaultseppunct}\relax
\EndOfBibitem
\bibitem[Wustrow and Neilson(2022)Wustrow, and
  Neilson]{wustrow_metathesis_2022}
Wustrow,~A.; Neilson,~J.~R. \emph{Reference {Module} in {Chemistry},
  {Molecular} {Sciences} and {Chemical} {Engineering}}; Elsevier, 2022; p
  B9780128231449000753\relax
\mciteBstWouldAddEndPuncttrue
\mciteSetBstMidEndSepPunct{\mcitedefaultmidpunct}
{\mcitedefaultendpunct}{\mcitedefaultseppunct}\relax
\EndOfBibitem
\bibitem[Yokokawa(1999)]{yokokawa_generalized_1999}
Yokokawa,~H. Generalized chemical potential diagram and its applications to
  chemical reactions at interfaces between dissimilar materials. \emph{Journal
  of Phase Equilibria} \textbf{1999}, \emph{20}, 258--287\relax
\mciteBstWouldAddEndPuncttrue
\mciteSetBstMidEndSepPunct{\mcitedefaultmidpunct}
{\mcitedefaultendpunct}{\mcitedefaultseppunct}\relax
\EndOfBibitem
\bibitem[Jain \latin{et~al.}(2013)Jain, Ong, Hautier, Chen, Richards, Dacek,
  Cholia, Gunter, Skinner, Ceder, and Persson]{jain_commentary_2013}
Jain,~A.; Ong,~S.~P.; Hautier,~G.; Chen,~W.; Richards,~W.~D.; Dacek,~S.;
  Cholia,~S.; Gunter,~D.; Skinner,~D.; Ceder,~G.; Persson,~K.~A. Commentary:
  {The} {Materials} {Project}: {A} materials genome approach to accelerating
  materials innovation. \emph{APL Materials} \textbf{2013}, \emph{1},
  011002\relax
\mciteBstWouldAddEndPuncttrue
\mciteSetBstMidEndSepPunct{\mcitedefaultmidpunct}
{\mcitedefaultendpunct}{\mcitedefaultseppunct}\relax
\EndOfBibitem
\bibitem[Saal \latin{et~al.}(2013)Saal, Kirklin, Aykol, Meredig, and
  Wolverton]{saal_materials_2013}
Saal,~J.~E.; Kirklin,~S.; Aykol,~M.; Meredig,~B.; Wolverton,~C. Materials
  {Design} and {Discovery} with {High}-{Throughput} {Density} {Functional}
  {Theory}: {The} {Open} {Quantum} {Materials} {Database} ({OQMD}). \emph{JOM}
  \textbf{2013}, \emph{65}, 1501--1509\relax
\mciteBstWouldAddEndPuncttrue
\mciteSetBstMidEndSepPunct{\mcitedefaultmidpunct}
{\mcitedefaultendpunct}{\mcitedefaultseppunct}\relax
\EndOfBibitem
\bibitem[Curtarolo \latin{et~al.}(2012)Curtarolo, Setyawan, Hart, Jahnatek,
  Chepulskii, Taylor, Wang, Xue, Yang, Levy, Mehl, Stokes, Demchenko, and
  Morgan]{curtarolo_aflow_2012}
Curtarolo,~S.; Setyawan,~W.; Hart,~G.~L.; Jahnatek,~M.; Chepulskii,~R.~V.;
  Taylor,~R.~H.; Wang,~S.; Xue,~J.; Yang,~K.; Levy,~O.; Mehl,~M.~J.;
  Stokes,~H.~T.; Demchenko,~D.~O.; Morgan,~D. {AFLOW}: {An} automatic framework
  for high-throughput materials discovery. \emph{Computational Materials
  Science} \textbf{2012}, \emph{58}, 218--226\relax
\mciteBstWouldAddEndPuncttrue
\mciteSetBstMidEndSepPunct{\mcitedefaultmidpunct}
{\mcitedefaultendpunct}{\mcitedefaultseppunct}\relax
\EndOfBibitem
\bibitem[Shoemaker \latin{et~al.}(2014)Shoemaker, Hu, Chung, Halder, Chupas,
  Soderholm, Mitchell, and Kanatzidis]{shoemaker_situ_2014}
Shoemaker,~D.~P.; Hu,~Y.-J.; Chung,~D.~Y.; Halder,~G.~J.; Chupas,~P.~J.;
  Soderholm,~L.; Mitchell,~J.~F.; Kanatzidis,~M.~G. In situ studies of a
  platform for metastable inorganic crystal growth and materials discovery.
  \emph{Proceedings of the National Academy of Sciences} \textbf{2014},
  \emph{111}, 10922--10927\relax
\mciteBstWouldAddEndPuncttrue
\mciteSetBstMidEndSepPunct{\mcitedefaultmidpunct}
{\mcitedefaultendpunct}{\mcitedefaultseppunct}\relax
\EndOfBibitem
\bibitem[Kovnir(2021)]{kovnir2021predictive}
Kovnir,~K. Predictive synthesis. \emph{Chemistry of Materials} \textbf{2021},
  \emph{33}, 4835--4841\relax
\mciteBstWouldAddEndPuncttrue
\mciteSetBstMidEndSepPunct{\mcitedefaultmidpunct}
{\mcitedefaultendpunct}{\mcitedefaultseppunct}\relax
\EndOfBibitem
\bibitem[Miura \latin{et~al.}(2021)Miura, Bartel, Goto, Mizuguchi, Moriyoshi,
  Kuroiwa, Wang, Yaguchi, Shirai, Nagao, Rosero‐Navarro, Tadanaga, Ceder, and
  Sun]{miura_observing_2021}
Miura,~A.; Bartel,~C.~J.; Goto,~Y.; Mizuguchi,~Y.; Moriyoshi,~C.; Kuroiwa,~Y.;
  Wang,~Y.; Yaguchi,~T.; Shirai,~M.; Nagao,~M.; Rosero‐Navarro,~N.~C.;
  Tadanaga,~K.; Ceder,~G.; Sun,~W. Observing and {Modeling} the {Sequential}
  {Pairwise} {Reactions} that {Drive} {Solid}‐{State} {Ceramic} {Synthesis}.
  \emph{Advanced Materials} \textbf{2021}, \emph{33}, 2100312\relax
\mciteBstWouldAddEndPuncttrue
\mciteSetBstMidEndSepPunct{\mcitedefaultmidpunct}
{\mcitedefaultendpunct}{\mcitedefaultseppunct}\relax
\EndOfBibitem
\bibitem[Richards \latin{et~al.}(2016)Richards, Miara, Wang, Kim, and
  Ceder]{richards_interface_2016}
Richards,~W.~D.; Miara,~L.~J.; Wang,~Y.; Kim,~J.~C.; Ceder,~G. Interface
  {Stability} in {Solid}-{State} {Batteries}. \emph{Chemistry of Materials}
  \textbf{2016}, \emph{28}, 266--273\relax
\mciteBstWouldAddEndPuncttrue
\mciteSetBstMidEndSepPunct{\mcitedefaultmidpunct}
{\mcitedefaultendpunct}{\mcitedefaultseppunct}\relax
\EndOfBibitem
\bibitem[Bianchini \latin{et~al.}(2020)Bianchini, Wang, Clément, Ouyang, Xiao,
  Kitchaev, Shi, Zhang, Wang, Kim, Zhang, Bai, Wang, Sun, and
  Ceder]{bianchini_interplay_2020}
Bianchini,~M.; Wang,~J.; Clément,~R.~J.; Ouyang,~B.; Xiao,~P.; Kitchaev,~D.;
  Shi,~T.; Zhang,~Y.; Wang,~Y.; Kim,~H.; Zhang,~M.; Bai,~J.; Wang,~F.; Sun,~W.;
  Ceder,~G. The interplay between thermodynamics and kinetics in the
  solid-state synthesis of layered oxides. \emph{Nature Materials}
  \textbf{2020}, \emph{19}, 1088--1095\relax
\mciteBstWouldAddEndPuncttrue
\mciteSetBstMidEndSepPunct{\mcitedefaultmidpunct}
{\mcitedefaultendpunct}{\mcitedefaultseppunct}\relax
\EndOfBibitem
\bibitem[Aykol \latin{et~al.}(2021)Aykol, Montoya, and
  Hummelshøj]{aykol_rational_2021}
Aykol,~M.; Montoya,~J.~H.; Hummelshøj,~J. Rational {Solid}-{State} {Synthesis}
  {Routes} for {Inorganic} {Materials}. \emph{Journal of the American Chemical
  Society} \textbf{2021}, \emph{143}, 9244--9259\relax
\mciteBstWouldAddEndPuncttrue
\mciteSetBstMidEndSepPunct{\mcitedefaultmidpunct}
{\mcitedefaultendpunct}{\mcitedefaultseppunct}\relax
\EndOfBibitem
\bibitem[Schmalzried(2008)]{schmalzried_chemical_2008}
Schmalzried,~H. \emph{Chemical {Kinetics} of {Solids}}, 1st ed.; Wiley-VCH:
  Weinheim, 2008; OCLC: 890041543\relax
\mciteBstWouldAddEndPuncttrue
\mciteSetBstMidEndSepPunct{\mcitedefaultmidpunct}
{\mcitedefaultendpunct}{\mcitedefaultseppunct}\relax
\EndOfBibitem
\bibitem[Johnson(1998)]{johnson_controlled_1998}
Johnson,~D.~C. Controlled synthesis of new compounds using modulated elemental
  reactants. \emph{Current Opinion in Solid State and Materials Science}
  \textbf{1998}, \emph{3}, 159--167\relax
\mciteBstWouldAddEndPuncttrue
\mciteSetBstMidEndSepPunct{\mcitedefaultmidpunct}
{\mcitedefaultendpunct}{\mcitedefaultseppunct}\relax
\EndOfBibitem
\bibitem[McAuliffe \latin{et~al.}(2022)McAuliffe, Huang, Montiel, Mehta, Davis,
  Petrova, Browning, Neilson, Liu, Thornton, and
  Veith]{mcauliffe_thin-film_2022}
McAuliffe,~R.~D.; Huang,~G.; Montiel,~D.; Mehta,~A.; Davis,~R.~C.; Petrova,~V.;
  Browning,~K.~L.; Neilson,~J.~R.; Liu,~P.; Thornton,~K.; Veith,~G.~M.
  Thin-{Film} {Paradigm} to {Probe} {Interfacial} {Diffusion} during
  {Solid}-{State} {Metathesis} {Reactions}. \emph{Chemistry of Materials}
  \textbf{2022}, \emph{34}, 6279--6287\relax
\mciteBstWouldAddEndPuncttrue
\mciteSetBstMidEndSepPunct{\mcitedefaultmidpunct}
{\mcitedefaultendpunct}{\mcitedefaultseppunct}\relax
\EndOfBibitem
\bibitem[Kamm \latin{et~al.}(2022)Kamm, Huang, Vornholt, McAuliffe, Veith,
  Thornton, and Chapman]{kamm_relative_2022}
Kamm,~G.~E.; Huang,~G.; Vornholt,~S.~M.; McAuliffe,~R.~D.; Veith,~G.~M.;
  Thornton,~K.~S.; Chapman,~K.~W. Relative {Kinetics} of {Solid}-{State}
  {Reactions}: {The} {Role} of {Architecture} in {Controlling} {Reactivity}.
  \emph{Journal of the American Chemical Society} \textbf{2022}, \emph{144},
  11975--11979\relax
\mciteBstWouldAddEndPuncttrue
\mciteSetBstMidEndSepPunct{\mcitedefaultmidpunct}
{\mcitedefaultendpunct}{\mcitedefaultseppunct}\relax
\EndOfBibitem
\bibitem[McDermott \latin{et~al.}(2021)McDermott, Dwaraknath, and
  Persson]{mcdermott_graph-based_2021}
McDermott,~M.~J.; Dwaraknath,~S.~S.; Persson,~K.~A. A graph-based network for
  predicting chemical reaction pathways in solid-state materials synthesis.
  \emph{Nature Communications} \textbf{2021}, \emph{12}, 3097\relax
\mciteBstWouldAddEndPuncttrue
\mciteSetBstMidEndSepPunct{\mcitedefaultmidpunct}
{\mcitedefaultendpunct}{\mcitedefaultseppunct}\relax
\EndOfBibitem
\bibitem[Walser and Bené(1976)Walser, and Bené]{walser_first_1976}
Walser,~R.~M.; Bené,~R.~W. First phase nucleation in
  silicon–transition‐metal planar interfaces. \emph{Applied Physics
  Letters} \textbf{1976}, \emph{28}, 624--625\relax
\mciteBstWouldAddEndPuncttrue
\mciteSetBstMidEndSepPunct{\mcitedefaultmidpunct}
{\mcitedefaultendpunct}{\mcitedefaultseppunct}\relax
\EndOfBibitem
\bibitem[Rom \latin{et~al.}(2021)Rom, Fallon, Wustrow, Prieto, and
  Neilson]{rom_bulk_2021}
Rom,~C.~L.; Fallon,~M.~J.; Wustrow,~A.; Prieto,~A.~L.; Neilson,~J.~R. Bulk
  {Synthesis}, {Structure}, and {Electronic} {Properties} of {Magnesium}
  {Zirconium} {Nitride} {Solid} {Solutions}. \emph{Chemistry of Materials}
  \textbf{2021}, \emph{33}, 5345--5354\relax
\mciteBstWouldAddEndPuncttrue
\mciteSetBstMidEndSepPunct{\mcitedefaultmidpunct}
{\mcitedefaultendpunct}{\mcitedefaultseppunct}\relax
\EndOfBibitem
\bibitem[d'Heurle(1988)]{dheurle_nucleation_1988}
d'Heurle,~F.~M. Nucleation of a new phase from the interaction of two adjacent
  phases: {Some} silicides. \emph{Journal of Materials Research} \textbf{1988},
  \emph{3}, 167--195\relax
\mciteBstWouldAddEndPuncttrue
\mciteSetBstMidEndSepPunct{\mcitedefaultmidpunct}
{\mcitedefaultendpunct}{\mcitedefaultseppunct}\relax
\EndOfBibitem
\bibitem[Wen \latin{et~al.}(2023)Wen, Spotte-Smith, Blau, McDermott,
  Krishnapriyan, and Persson]{wen_chemical_2023}
Wen,~M.; Spotte-Smith,~E. W.~C.; Blau,~S.~M.; McDermott,~M.~J.;
  Krishnapriyan,~A.~S.; Persson,~K.~A. Chemical reaction networks and
  opportunities for machine learning. \emph{Nature Computational Science}
  \textbf{2023}, \relax
\mciteBstWouldAddEndPunctfalse
\mciteSetBstMidEndSepPunct{\mcitedefaultmidpunct}
{}{\mcitedefaultseppunct}\relax
\EndOfBibitem
\bibitem[Toberer \latin{et~al.}(2004)Toberer, Weaver, Ramesha, and
  Seshadri]{Toberer2004}
Toberer,~E.~S.; Weaver,~J.~C.; Ramesha,~K.; Seshadri,~R. Macroporous monoliths
  of functional perovskite materials through assisted metathesis.
  \emph{Chemistry of Materials} \textbf{2004}, \emph{16}, 2194--2200\relax
\mciteBstWouldAddEndPuncttrue
\mciteSetBstMidEndSepPunct{\mcitedefaultmidpunct}
{\mcitedefaultendpunct}{\mcitedefaultseppunct}\relax
\EndOfBibitem
\bibitem[Todd and Neilson(2019)Todd, and Neilson]{todd_selective_2019}
Todd,~P.~K.; Neilson,~J.~R. Selective {Formation} of {Yttrium} {Manganese}
  {Oxides} through {Kinetically} {Competent} {Assisted} {Metathesis}
  {Reactions}. \emph{Journal of the American Chemical Society} \textbf{2019},
  \emph{141}, 1191--1195\relax
\mciteBstWouldAddEndPuncttrue
\mciteSetBstMidEndSepPunct{\mcitedefaultmidpunct}
{\mcitedefaultendpunct}{\mcitedefaultseppunct}\relax
\EndOfBibitem
\bibitem[Fedorova \latin{et~al.}(2004)Fedorova, Titova, Balakirev, and
  Golikov]{fedorova_heterogeneous_2004}
Fedorova,~O.~M.; Titova,~S.~G.; Balakirev,~V.~F.; Golikov,~Y.~V. Heterogeneous
  equilibria in the {Y}-{Mn}-{O} and {Tm}-{Mn}-{O} systems in air.
  \emph{Russian Journal of Applied Chemistry} \textbf{2004}, \emph{77},
  1556--1558\relax
\mciteBstWouldAddEndPuncttrue
\mciteSetBstMidEndSepPunct{\mcitedefaultmidpunct}
{\mcitedefaultendpunct}{\mcitedefaultseppunct}\relax
\EndOfBibitem
\bibitem[Smith \latin{et~al.}(2009)Smith, Mizoguchi, Delaney, Spaldin, Sleight,
  and Subramanian]{smith_mn_2009}
Smith,~A.~E.; Mizoguchi,~H.; Delaney,~K.; Spaldin,~N.~A.; Sleight,~A.~W.;
  Subramanian,~M.~A. Mn $^{\textrm{3+}}$ in {Trigonal} {Bipyramidal}
  {Coordination}: {A} {New} {Blue} {Chromophore}. \emph{Journal of the American
  Chemical Society} \textbf{2009}, \emph{131}, 17084--17086\relax
\mciteBstWouldAddEndPuncttrue
\mciteSetBstMidEndSepPunct{\mcitedefaultmidpunct}
{\mcitedefaultendpunct}{\mcitedefaultseppunct}\relax
\EndOfBibitem
\bibitem[Wood \latin{et~al.}(1973)Wood, Austin, Collings, and
  Brog]{wood_magnetic_1973}
Wood,~V.~E.; Austin,~A.~E.; Collings,~E.~W.; Brog,~K.~C. Magnetic properties of
  heavy-rare-earth orthomanganites. \emph{Journal of Physics and Chemistry of
  Solids} \textbf{1973}, \emph{34}, 859--868\relax
\mciteBstWouldAddEndPuncttrue
\mciteSetBstMidEndSepPunct{\mcitedefaultmidpunct}
{\mcitedefaultendpunct}{\mcitedefaultseppunct}\relax
\EndOfBibitem
\bibitem[Brinks \latin{et~al.}(1997)Brinks, Fjellvåg, and
  Kjekshus]{brinks_synthesis_1997}
Brinks,~H.; Fjellvåg,~H.; Kjekshus,~A. Synthesis of {Metastable}
  {Perovskite}-type {YMnO3} and {HoMnO3}. \emph{Journal of Solid State
  Chemistry} \textbf{1997}, \emph{129}, 334--340\relax
\mciteBstWouldAddEndPuncttrue
\mciteSetBstMidEndSepPunct{\mcitedefaultmidpunct}
{\mcitedefaultendpunct}{\mcitedefaultseppunct}\relax
\EndOfBibitem
\bibitem[Todd \latin{et~al.}(2020)Todd, Wustrow, McAuliffe, McDermott, Tran,
  McBride, Boeding, O’Nolan, Liu, Dwaraknath, Chapman, Billinge, Persson,
  Huq, Veith, and Neilson]{todd_defect-accommodating_2020}
Todd,~P.~K. \latin{et~al.}  Defect-{Accommodating} {Intermediates} {Yield}
  {Selective} {Low}-{Temperature} {Synthesis} of {YMnO} $_{\textrm{3}}$
  {Polymorphs}. \emph{Inorganic Chemistry} \textbf{2020}, \emph{59},
  13639--13650\relax
\mciteBstWouldAddEndPuncttrue
\mciteSetBstMidEndSepPunct{\mcitedefaultmidpunct}
{\mcitedefaultendpunct}{\mcitedefaultseppunct}\relax
\EndOfBibitem
\bibitem[Todd \latin{et~al.}(2019)Todd, Smith, and Neilson]{todd_yttrium_2019}
Todd,~P.~K.; Smith,~A. M.~M.; Neilson,~J.~R. Yttrium {Manganese} {Oxide}
  {Phase} {Stability} and {Selectivity} {Using} {Lithium} {Carbonate}
  {Assisted} {Metathesis} {Reactions}. \emph{Inorganic Chemistry}
  \textbf{2019}, \emph{58}, 15166--15174\relax
\mciteBstWouldAddEndPuncttrue
\mciteSetBstMidEndSepPunct{\mcitedefaultmidpunct}
{\mcitedefaultendpunct}{\mcitedefaultseppunct}\relax
\EndOfBibitem
\bibitem[Fujinaka \latin{et~al.}(1979)Fujinaka, Kinomura, Koizumi, Miyamoto,
  and Kume]{fujinaka_syntheses_1979}
Fujinaka,~H.; Kinomura,~N.; Koizumi,~M.; Miyamoto,~Y.; Kume,~S. Syntheses and
  physical properties of pyrochlore-type {A2B2O7} ({A}={Tl},{Y};
  {B}={Cr},{Mn}). \emph{Materials Research Bulletin} \textbf{1979}, \emph{14},
  1133--1137, ISBN: 0911330550\relax
\mciteBstWouldAddEndPuncttrue
\mciteSetBstMidEndSepPunct{\mcitedefaultmidpunct}
{\mcitedefaultendpunct}{\mcitedefaultseppunct}\relax
\EndOfBibitem
\bibitem[Subramanian \latin{et~al.}(1988)Subramanian, Torardi, Johnson,
  Pannetier, and Sleight]{subramanian_ferromagnetic_1988}
Subramanian,~M.~A.; Torardi,~C.~C.; Johnson,~D.~C.; Pannetier,~J.;
  Sleight,~A.~W. Ferromagnetic {R2Mn207} {Pyrochlores} ( {R} = {Dy}-{Lu} , {Y}
  )*. \textbf{1988}, \emph{30}, 24--30\relax
\mciteBstWouldAddEndPuncttrue
\mciteSetBstMidEndSepPunct{\mcitedefaultmidpunct}
{\mcitedefaultendpunct}{\mcitedefaultseppunct}\relax
\EndOfBibitem
\bibitem[Bartel \latin{et~al.}(2018)Bartel, Millican, Deml, Rumptz, Tumas,
  Weimer, Lany, Stevanović, Musgrave, and Holder]{bartel_physical_2018}
Bartel,~C.~J.; Millican,~S.~L.; Deml,~A.~M.; Rumptz,~J.~R.; Tumas,~W.;
  Weimer,~A.~W.; Lany,~S.; Stevanović,~V.; Musgrave,~C.~B.; Holder,~A.~M.
  Physical descriptor for the {Gibbs} energy of inorganic crystalline solids
  and temperature-dependent materials chemistry. \emph{Nature Communications}
  \textbf{2018}, \emph{9}, 4168\relax
\mciteBstWouldAddEndPuncttrue
\mciteSetBstMidEndSepPunct{\mcitedefaultmidpunct}
{\mcitedefaultendpunct}{\mcitedefaultseppunct}\relax
\EndOfBibitem
\bibitem[Todd \latin{et~al.}(2021)Todd, McDermott, Rom, Corrao, Denney,
  Dwaraknath, Khalifah, Persson, and Neilson]{todd_selectivity_2021}
Todd,~P.~K.; McDermott,~M.~J.; Rom,~C.~L.; Corrao,~A.~A.; Denney,~J.~J.;
  Dwaraknath,~S.~S.; Khalifah,~P.~G.; Persson,~K.~A.; Neilson,~J.~R.
  Selectivity in {Yttrium} {Manganese} {Oxide} {Synthesis} via {Local}
  {Chemical} {Potentials} in {Hyperdimensional} {Phase} {Space}. \emph{Journal
  of the American Chemical Society} \textbf{2021}, \emph{143},
  15185--15194\relax
\mciteBstWouldAddEndPuncttrue
\mciteSetBstMidEndSepPunct{\mcitedefaultmidpunct}
{\mcitedefaultendpunct}{\mcitedefaultseppunct}\relax
\EndOfBibitem
\bibitem[Martinolich and Neilson(2017)Martinolich, and
  Neilson]{martinolich_toward_2017}
Martinolich,~A.~J.; Neilson,~J.~R. Toward {Reaction}-by-{Design}: {Achieving}
  {Kinetic} {Control} of {Solid} {State} {Chemistry} with {Metathesis}.
  \emph{Chemistry of Materials} \textbf{2017}, \emph{29}, 479--489\relax
\mciteBstWouldAddEndPuncttrue
\mciteSetBstMidEndSepPunct{\mcitedefaultmidpunct}
{\mcitedefaultendpunct}{\mcitedefaultseppunct}\relax
\EndOfBibitem
\bibitem[Giaquinta and zur Loye(1994)Giaquinta, and zur
  Loye]{giaquinta_structural_1994}
Giaquinta,~D.~M.; zur Loye,~H.-C. Structural {Predictions} in the {ABO3}
  {Phase} {Diagram}. \emph{Chemistry of Materials} \textbf{1994}, \emph{6},
  365--372, ISBN: 0897-4756\relax
\mciteBstWouldAddEndPuncttrue
\mciteSetBstMidEndSepPunct{\mcitedefaultmidpunct}
{\mcitedefaultendpunct}{\mcitedefaultseppunct}\relax
\EndOfBibitem
\bibitem[Sun and Powell-Palm(2021)Sun, and Powell-Palm]{sun_generalized_2021}
Sun,~W.; Powell-Palm,~M.~J. Generalized {Gibbs}' {Phase} {Rule}. \textbf{2021},
  Publisher: arXiv Version Number: 1\relax
\mciteBstWouldAddEndPuncttrue
\mciteSetBstMidEndSepPunct{\mcitedefaultmidpunct}
{\mcitedefaultendpunct}{\mcitedefaultseppunct}\relax
\EndOfBibitem
\bibitem[Kononova \latin{et~al.}(2019)Kononova, Huo, He, Rong, Botari, Sun,
  Tshitoyan, and Ceder]{kononova_text-mined_2019}
Kononova,~O.; Huo,~H.; He,~T.; Rong,~Z.; Botari,~T.; Sun,~W.; Tshitoyan,~V.;
  Ceder,~G. Text-mined dataset of inorganic materials synthesis recipes.
  \emph{Scientific Data} \textbf{2019}, \emph{6}, 203\relax
\mciteBstWouldAddEndPuncttrue
\mciteSetBstMidEndSepPunct{\mcitedefaultmidpunct}
{\mcitedefaultendpunct}{\mcitedefaultseppunct}\relax
\EndOfBibitem
\bibitem[Kim \latin{et~al.}(2017)Kim, Huang, Saunders, McCallum, Ceder, and
  Olivetti]{kim_materials_2017}
Kim,~E.; Huang,~K.; Saunders,~A.; McCallum,~A.; Ceder,~G.; Olivetti,~E.
  Materials {Synthesis} {Insights} from {Scientific} {Literature} via {Text}
  {Extraction} and {Machine} {Learning}. \emph{Chemistry of Materials}
  \textbf{2017}, \emph{29}, 9436--9444\relax
\mciteBstWouldAddEndPuncttrue
\mciteSetBstMidEndSepPunct{\mcitedefaultmidpunct}
{\mcitedefaultendpunct}{\mcitedefaultseppunct}\relax
\EndOfBibitem
\bibitem[He \latin{et~al.}(2020)He, Sun, Huo, Kononova, Rong, Tshitoyan,
  Botari, and Ceder]{he_similarity_2020}
He,~T.; Sun,~W.; Huo,~H.; Kononova,~O.; Rong,~Z.; Tshitoyan,~V.; Botari,~T.;
  Ceder,~G. Similarity of {Precursors} in {Solid}-{State} {Synthesis} as
  {Text}-{Mined} from {Scientific} {Literature}. \emph{Chemistry of Materials}
  \textbf{2020}, \emph{32}, 7861--7873\relax
\mciteBstWouldAddEndPuncttrue
\mciteSetBstMidEndSepPunct{\mcitedefaultmidpunct}
{\mcitedefaultendpunct}{\mcitedefaultseppunct}\relax
\EndOfBibitem
\bibitem[Huo \latin{et~al.}(2022)Huo, Bartel, He, Trewartha, Dunn, Ouyang,
  Jain, and Ceder]{huo_machine-learning_2022}
Huo,~H.; Bartel,~C.~J.; He,~T.; Trewartha,~A.; Dunn,~A.; Ouyang,~B.; Jain,~A.;
  Ceder,~G. Machine-learning rationalization and prediction of solid-state
  synthesis conditions. \emph{Chemistry of Materials} \textbf{2022}, \emph{34},
  7323--7336, arXiv:2204.08151 [cond-mat]\relax
\mciteBstWouldAddEndPuncttrue
\mciteSetBstMidEndSepPunct{\mcitedefaultmidpunct}
{\mcitedefaultendpunct}{\mcitedefaultseppunct}\relax
\EndOfBibitem
\bibitem[Phelan \latin{et~al.}(2013)Phelan, Wallace, Arpino, Neilson, Livi,
  Seabourne, Scott, and McQueen]{phelan_stacking_2013}
Phelan,~W.~A.; Wallace,~D.~C.; Arpino,~K.~E.; Neilson,~J.~R.; Livi,~K.~J.;
  Seabourne,~C.~R.; Scott,~A.~J.; McQueen,~T.~M. Stacking {Variants} and
  {Superconductivity} in the {Bi}–{O}–{S} {System}. \emph{Journal of the
  American Chemical Society} \textbf{2013}, \emph{135}, 5372--5374\relax
\mciteBstWouldAddEndPuncttrue
\mciteSetBstMidEndSepPunct{\mcitedefaultmidpunct}
{\mcitedefaultendpunct}{\mcitedefaultseppunct}\relax
\EndOfBibitem
\bibitem[Wustrow \latin{et~al.}(2022)Wustrow, McDermott, O’Nolan, Liu, Tran,
  McBride, Vornholt, Feng, Dwaraknath, Chapman, Billinge, Sun, Persson, and
  Neilson]{wustrow_reaction_2022}
Wustrow,~A.; McDermott,~M.~J.; O’Nolan,~D.; Liu,~C.-H.; Tran,~G.~T.;
  McBride,~B.~C.; Vornholt,~S.~M.; Feng,~C.; Dwaraknath,~S.~S.; Chapman,~K.~W.;
  Billinge,~S. J.~L.; Sun,~W.; Persson,~K.~A.; Neilson,~J.~R. Reaction
  {Selectivity} in {Cometathesis}: {Yttrium} {Manganese} {Oxides}.
  \emph{Chemistry of Materials} \textbf{2022}, \emph{34}, 4694--4702\relax
\mciteBstWouldAddEndPuncttrue
\mciteSetBstMidEndSepPunct{\mcitedefaultmidpunct}
{\mcitedefaultendpunct}{\mcitedefaultseppunct}\relax
\EndOfBibitem
\bibitem[Dybkov(1986)]{dybkov_reaction_1986}
Dybkov,~V.~I. Reaction diffusion in heterogeneous binary systems: {Part} 1
  {Growth} of the chemical compound layers at the interface between two
  elementary substances: one compound layer. \emph{Journal of Materials
  Science} \textbf{1986}, \emph{21}, 3078--3084\relax
\mciteBstWouldAddEndPuncttrue
\mciteSetBstMidEndSepPunct{\mcitedefaultmidpunct}
{\mcitedefaultendpunct}{\mcitedefaultseppunct}\relax
\EndOfBibitem
\end{mcitethebibliography}

\end{document}